\documentclass [preprint, prd, aps, nofootinbib]{revtex4}

\usepackage{graphicx}

\begin{document}
\draft
\title{\large \bf Probing spacetime foam with
extragalactic sources}

\author{W. A. Christiansen,
Y. Jack Ng\footnote{Corresponding author.  Electronic address:
yjng@physics.unc.edu}
and H. van Dam}
\address{Department
of Physics and Astronomy, University of North Carolina, Chapel
Hill, NC 27599, USA\\}

\bigskip

\begin{abstract}

Due to quantum fluctuations, spacetime is probably
``foamy'' on very small 
scales.
We propose to detect this texture of spacetime foam by 
looking for halo structures in
the images of distant quasars.  We find that
the Very Large Telescope interferometer
will be on the verge of being able to probe the fabric of spacetime
when it reaches its design performance.  
Our method also allows us to
use spacetime foam physics and physics of computation to 
infer the existence of dark energy/matter,
independent of the evidence from recent cosmological
observations.

\bigskip



\end{abstract}
\maketitle


Before the last century, spacetime was regarded as nothing
more than a passive and static arena in which events took place.
Early last century, Einstein's general relativity
changed that viewpoint, envisioning spacetime as an
active and dynamical entity.  Nowadays many physicists also believe
that spacetime, like all matter and energy, undergoes quantum
fluctuations.  These quantum fluctuations make spacetime
foamy on small spacetime scales.\cite{Wheeler,
DeWitt,Hawking,Ashtekar,Ellis,ng03b}
But how large are the fluctuations?  How foamy is spacetime?
There is no consensus yet on the answers to these questions.\cite{ng03b}
But we do expect the foaminess of spacetime to lead to uncertainties in
distance ($l$) measurements given by
$\delta l \gtrsim l^{1 - \alpha} l_P^{\alpha}$, where $l_P \equiv
(\hbar G/c^3)^{1/2} \sim 10^{-33}$cm is the minuscule Planck length, the 
intrinsic length scale characterizing quantum gravity and hence relevant to
spacetime foam physics.  The parameter
$\alpha \sim 1$ specifies different spacetime foam models.
Because the Planck length is so tiny, spacetime foam effects
are naturally exceedingly small.  The trick will be to find ways
to amplify the minuscule effects.\cite{AC}  One way is
to accumulate effects of spacetime fluctuations over a huge distance.
Here we propose
using spacetime foam-induced fluctuations in the
direction of the wave vector of light from extragalactic point sources to
detect spacetime foam.  Intriguingly this method will take us 
within striking distance of probing the fabric of spacetime once 
the Very Large Telescope (VLT) 
interferometer reaches its design performance.
Furthermore, as shown below, this method allows us to
infer the existence of unconventional (dark) matter or
energy\cite{Turner}, 
independent of the evidence from recent
supernovae observations\cite{SNa,SNa1}, observations on the cosmic
microwave background anisotropies\cite{WMAP}, and recent studies based on
local X-ray luminosity of galaxy clusters\cite{cluster} etc.


Although our discussion is applicable to all the spacetime foam models, for
concreteness, we will concentrate on two models corresponding to 
$\alpha = 2/3, 1/2$. 
The choice of $\alpha = 2/3$ 
\cite{ng94,ng95,kar66} yields a model that turns out 
to be consistent\cite{ng01} with black hole physics and
the holographic principle\citep{tho93, sus95} which, many theoretical 
physicists believe, correctly governs how densely information can
be packed in a region of space in its stipulation that the 
information contents can be encoded on the two-dimensional
surface around the region, like a hologram.
For good reasons, it has come to be known as the holographic 
model.\cite{ng03b}
The case $\alpha = 1/2$ corresponds to the random-walk model\cite{dio89,
ame99}, so called because the associated dependence on $\sqrt{l}$ is the
hallmark of fluctuations of a random-walk type.
There are various ways to derive the two models,\cite{giovanni} but here we
follow a recent argument \cite{llo04, gio04} used to
map out the geometry of spacetime.
The point is that quantum fluctuations of
spacetime manifest themselves in the form of uncertainties in the
geometry of spacetime.  Therefore, the structure
of spacetime foam can be inferred from the accuracy with which we can
measure that geometry.

One way to map out the geometry of spacetime
is to fill space with clocks,
exchanging signals with the other clocks and measuring the signals'
times of arrival.  (This is how the global positioning system works.)
We can think
of this procedure as a special type of computation,
one whose purpose
is to map out the geometry of spacetime.
So imagine using
a collection of clocks of total mass M to
map out a volume of radius $l$ over the amount time $T=l/c$ it
takes light to cross the volume.
The total number of elementary
events, including the ticks of the clocks and the measurements of signals,
that can take place within the volume over this time is constrained by
the Margolus-Levitin theorem\cite{mar98,GPP} in 
quantum computation, that bounds
the number of elementary logical operations that can be
performed within the volume.  Thus
the maximum number of ticks
and measurements is given by the
energy $E=Mc^2$, times the time $T$,
divided by Planck's constant.  To prevent black hole formation,
the total mass $M$ of clocks within a volume of radius $l$
must be less than $lc^2/2G$.  Together
these two limits imply that the total number of elementary events
(or the number of operations)
that can occur in the volume of spacetime is no greater than
$lTc/l_P^2 = l^2/l_P^2$.  Thus, the total number of ``cells'' in the
volume of spacetime is bounded by the area of the spatial part
of the volume.  In this Letter, we will ignore 
multiplicative constants of order 1, content with order-of-magnitude
estimates.

To maximize spatial resolution we can have each clock tick only once in
time $T$.  (That is, it takes the clock in each cell the same amount of
time to tick as it takes light to travel around the volume.)
Then each cell in this volume of radius $l$ occupies a space of
$l^3 /(l^2/l_P^2) = l l_P^2$, i.e., the cells are separated
by an average distance
of $l^{1/3} l_P^{2/3}$.
We interpret this result to mean that the uncertainty
$\delta l$ in the measurement of any distance $l$ cannot be smaller than
$l^{1/3} l_P^{2/3}$ on the average.\cite{average}  This yields the
holographic model of distance fluctuations.

On the other hand, if we spread the cells out uniformly in both space and
time, then
the spatial size
of each cell is $l^{1/2} l_P^{1/2}$, and the temporal separation of
successive ticks of each clock is $l^{1/2} l_P^{1/2}/c$ which is just
the time it takes a cell to communicate with a neighboring cell.
(As we will show below, this is the
accuracy with which ordinary matter maps out spacetime.  For later use,
we note that, for $l \sim 10^{28}$ cm, the size of the observable universe,
the average cell size is 
about $10^{-3}$ cm and it takes a cell about $10^{-14}$ sec to communicate
with a neighboring cell.)
This corresponds to the random-walk
model of spacetime fluctuations for which $\delta l \sim l^{1/2}
l_P^{1/2}$.

To probe spacetime foam with extragalactic sources, we first recall
that the phase of an electromagnetic wave
through spacetime foam is $kx - \omega t \pm \Delta \phi$,
where $\Delta \phi$ is random uncertainty introduced by spacetime
fluctuations.  This cumulative
phase uncertainty is a function of the distance
$l$ to the source given by \cite{ng03a}
$\Delta \phi \sim 2 \pi \Sigma (\delta \lambda / \lambda)
\sim 2 \pi \delta l / \lambda
\sim 2 \pi (l_P/\lambda)^{\alpha} (l/\lambda)^{1-\alpha}$,
where $\lambda$ is the wavelength of the observed light from the source,
and we have used the proper cumulative factor\cite{ng03a} 
$(l/\lambda)^{1 - \alpha}$ in the summation $\Sigma$
(over the $l/\lambda$
intervals each of length $\lambda$, forming the total distance $l$) 
to yield 
$\Sigma \delta \lambda = (l/ \lambda)^{1 - \alpha} \delta \lambda
\sim (l/ \lambda)^{1 - \alpha} \lambda^{1 - \alpha} l_P^{\alpha}
\sim \delta l$.
Thus there is a possibility\cite{lie03,rag03}
for detecting the Planck-level fluctuations predicted
by the various theoretical models (parametrized by $\alpha$) by
using interferometric techniques to search for fringes from point
sources in distant objects because of the amplification provided by the
factor $(l/\lambda)^{1-\alpha}$.

As discussed in more detail below,
when $\Delta \phi \sim \pi$ (unless noted otherwise, we measure angles
in radians), the cumulative uncertainty in the wave's
phase will have effectively scrambled the wave front sufficiently to
prevent the observation of interferometric fringes.
It should be noted, however, that intrinsic structure within galaxies (e.g.,
galaxies in the Hubble Deep Field) can mask spacetime foam effects, thus 
requiring that test targets be distant point sources 
such as quasars or ultra-bright AGN.
Let us therefore consider the case
of PKS1413+135 \cite{per02}, an AGN for which the redshift is $z = 0.2467$.
With $l \approx 1.2$ Gpc \cite{Handq} and $\lambda = 1.6 \mu$m,
we \cite{ng03a} find $\Delta \phi \sim 10 \times 2 \pi$ and
$10^{-9} \times 2 \pi$ for the random-walk model and
the holographic model of spacetime foam respectively.
Thus the observation \cite{per02}
by the Hubble Space Telescope (HST) of an Airy ring for PKS1413+135
marginally rules
out the random-walk model, but fails by $9$ orders of magnitude to test
the holographic model.  See also Ref.\cite{cou03}.

\begin{figure}
\centering
\includegraphics[height=4in]{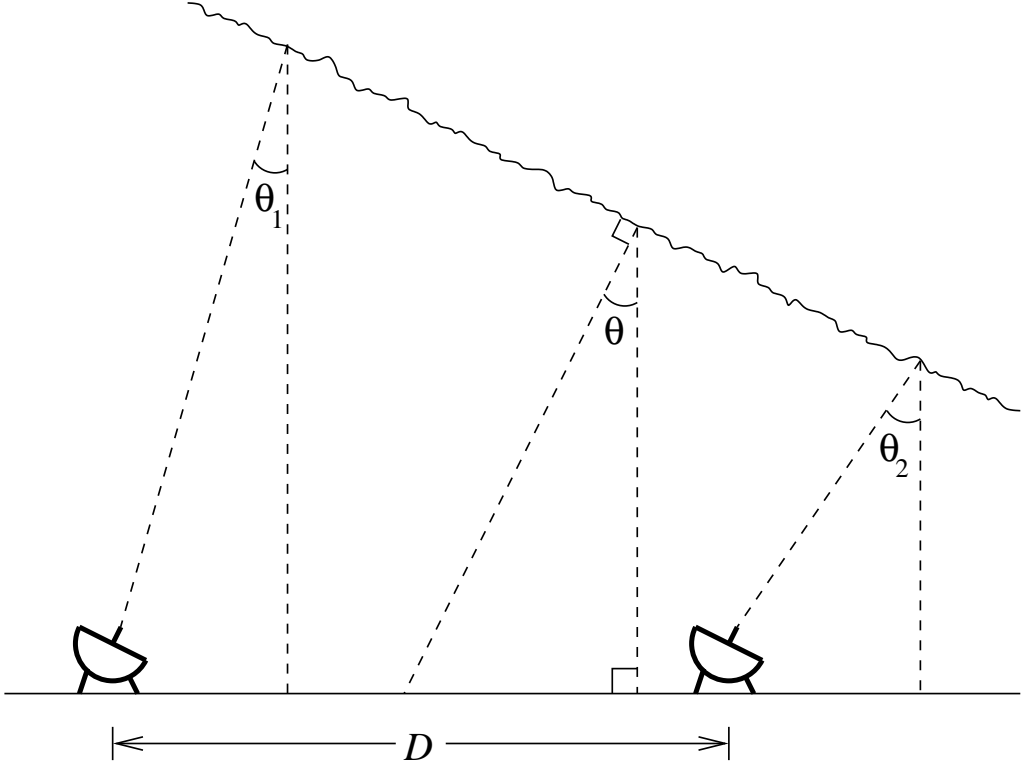}
\caption{
Interferometer observing an incoming electromagnetic wave from 
a distant galaxy.
The local wave
vector makes an angle $\theta$ with respect to the normal to the
interferometer baseline.
The tiny corrugations (greatly exaggerated in the Figure) 
in the wave front are due to 
spacetime foam-induced fluctuations in phase velocity.}
\label{fig1}
\end{figure}

However, it may be possible to test spacetime foam models more
stringently
than implied above by taking into account the expected scattering from
spacetime foam fluctuations, which is directly correlated with phase
fluctuations.  Consider a two-element interferometer observing an
incoming electromagnetic wave whose local wave vector $k$ makes an
angle $\theta$ with respect to the normal to the interferometer
baseline as shown in Fig.~\ref{fig1}.

Because of fluctuations in phase velocity, the wave front
develops tiny corrugations in which one portion of the wave front is
advanced while another is retarded.  The result is that the wave vector
acquires a cumulative random uncertainty in direction with an angular
spread of the order\cite{explanation} of
$\Sigma \frac{\delta k}{k} = \Sigma
\frac{\delta \lambda}{\lambda} \sim \frac{\delta l}{\lambda}
\sim {\Delta \phi \over 2 \pi}$, 
where we recall that $\Delta \phi \sim 2 \pi \delta l / \lambda$. 
Thus if the incoming wave has
an uncertainty in
\emph{both} its phase \emph{and} wave vector direction, the correlated
electric field from the two elements is
$E = E_0 \exp(i \psi_2 /2) + E_0 \exp(-i \psi_1 /2)$,
where
$\psi_{1(2)} = (2 \pi D/\lambda) \sin \theta_{1(2)} \pm \Delta \phi$,
with D denoting the interferometer baseline length,
and $\theta_{1(2)} \sim \theta \pm$ (fluctuations in wave vector direction).
Since the fluctuations in wave vector direction are of order $\Delta
\phi/ 2 \pi$, arbitrarily setting $\theta_2 = \theta$ and $\theta_1 = \theta
- \Delta \phi / 2 \pi$, we get, for the magnitude of the correlated electric 
field, 
\begin{equation}
|E| \simeq 2 E_0 
\left | \cos \left( {\pi \over 2} [2 \theta - \Delta \phi/2 \pi] 
{D \over \lambda} \right) \right |, 
\label{field2}
\end{equation}
where we have used the fact that $\theta$, $\Delta \phi$ and $\lambda / D$
are all much less than 1.

For $\Delta \phi = 0$, i.e., non-existent spacetime foam, the first
null occurs when $2 \theta = \lambda /D$.  But for
$\Delta \phi \neq 0$ due to spacetime foam effects,  
the first null for the halo will occur when $2 \theta 
\sim \lambda / D \pm \Delta \phi/ 2 \pi$. 
Since $\Delta \phi$ is a random phase fluctuation, the Michelson 
fringe visibility, $V = (P_{max} - P_{min})/(P_{max} + P_{min})$,
calculated from the complex square of Eq. (\ref{field2}) will
approximate the normalized visibility of a core-halo structure with
a halo angular diameter $2 \pi \lambda /D$.  That is,
\begin{equation}
V \left (\Delta \phi, {D \over \lambda}\right )
= f_c + (1 - f_c) {J_1\left({\Delta \phi \over 2 \pi}
{D \over \lambda} \right ) \over {\Delta \phi \over 2 \pi}
{D \over \lambda}},
\label{visibility}
\end{equation}
where $J_1$ is the first order Bessel function of the first kind
and we have introduced the parameter $f_c \ll 1$ to account for the 
possibility of bias in the spacetime foam scattering which may 
favor the forward direction.
Thus, there is 
a strong reduction in visibility when the argument of $J_1$ is of order
unity, i.e., $\Delta \phi \sim 2 \pi \lambda /D$, and  
spacetime foam theories can
be tested by imaging and interferometry for values of
$\Delta \phi$ which are orders of magnitude less than tests
based simply on the uncertainty in phase alone!

Again consider the case of PKS1413+135.
With $D = 2.4$ m for HST, we expect to detect halos
if $\Delta \phi \sim 10^{-6} \times 2 \pi$ (as compared to 
$\Delta \phi \sim 10^{-9} \times 2 \pi$).
Thus, the HST image only fails to test the holographic model by
approximately 3 orders of magnitude, rather than the $9$ orders
of magnitude discussed above for phase scrambling.
However,the absence of a spacetime foam induced halo structure in the 
HST image of PKS1413+135 rules
out convincingly the random-walk model.  In fact, the scaling relation
discussed above indicates
that all spacetime foam models with $\alpha \lesssim 0.6$ are ruled 
out by this HST observation.
We note that the case of a Hubble Deep
Field galaxy at z=5.34 considered in \cite{rag03} (for which
$l \sim 7.7$ Gpc and $\lambda \sim 814$ nm) gives similar bounds
(within a factor of $4$).  At $2 \mu$m, however, with the maximum
VLTI baseline of $140$m, the visibility of fringes from PKS$1413+135$ 
may be reduced by half because of spacetime foam scattering.

The key for detecting spacetime foam via interferometers, however, 
relates to the issues of
sensitivity and masking.  Sensitivity is simply the minimum
flux for which interferometer fringes
can be detected, $S_{min} = L/ 4 \pi l^2$, where
$L$ is the spectral luminosity of the source in the observational
band and $l$ is the distance to the source.  For a bright AGN with
a spectral luminosity in the near IR of $L_{32}$ (in units of 
$10^{32}$ ergs/s $\cdot$ Hz) and a sensitivity $S_{10}$ (in units of
$10$ mJy $\equiv 10^{-25}$ ergs/cm$^2 \cdot$ s $\cdot$ Hz),
the maximum distance that can be probed for spacetime
fluctuations is $l_{max} \sim 2 \times 10^9 \sqrt{L_{32}/S_{10}}$ (pc).

Masking is any structure of physical size $d_{source}$ 
that is resolved, and, 
therefore, whose angular size $\beta = d_{source} / l \gtrsim
\Delta \phi$.  Note that since $\beta$ goes as $l^{-1}$ while 
$\Delta \phi$ goes as $l^{1/3}$ (for $\alpha = 2/3$), large distances
are required to avoid masking.  For example, at optical-IR wavelengths
the Broad Emission Line Region (BELR) of a typical AGN has a 
physical size $d_{BELR}$ between $0.1$ and 1 pc.  As a result, 
avoidance of possible masking (in the optical-IR band) indicates
that observations should concentrate on AGN farther than $
l_{min} \sim 1.7 \times 10^8$ parsecs (assuming the upper limit
of 1pc for $d_{BELR}$).  The above arguments then restrict tests to 
objects such as distant quasar/AGN cores which are highly luminous 
yet unresolved.

We note in passing that even though the 
Very Long Baseline Array has superior
resolution and sensibility compared to the VLTI, it is not suitable for 
searching for spacetime foam scattering because the effects
of the quantum foam are inversely proportional to wavelength which
would therefore require a resolution $\sim 4$ orders of 
magnitude finer at radio wavelengths.  At the opposite wavelength 
extreme the shorter X-ray wavelength ($\sim 10^{-8}$ cm) of the Chandra 
X-ray Observatory might make up for the lack of angular
resolution ($\sim 1$ arcsec), provided X-ray quasars can be found which
are sufficiently bright to allow for corrections for the broad wings of 
the Chandra point spread function.  We shall explore this possibility in
a follow up paper.

The fact that the random-walk model is ruled out has an interesting
cosmological implication as we will now show.  But first we have
to discuss the information contents of the universe.
Merely by existing,
all physical systems register information.\cite{llo02, llo04}
And by evolving
dynamically in time, they transform and process that information.
In other words, they compute.  By extrapolation, the whole universe
can be regarded as a computer.

We know that our universe is
close to its critical density, about $10^{-9}$ joule per cubic meter,
and the present Hubble
horizon of the observable universe is about $10^{28}$ cm.
Using the Margolus-Levitin theorem, one finds that the universe
can perform up to $10^{106}$ operations per second.  And just as
in the discussion of limits to mapping the geometry of spacetime
above, the maximum number of elementary operations that
can be performed by our universe
in space of the Hubble size $R_H$ over time $T \sim R_H /c$
without undergoing gravitational
collapse is just $R_H^2/l_P^2$.  Therefore our universe can
have performed no more than $10^{123}$ ops.

We can now use statistical mechanics to
calculate the total amount of information that could be stored on
conventional matter,\cite{llo02, llo04} such as atoms and photons, 
within the horizon.
Matter can embody the maximum information when it is converted
to energetic and effectively massless particles.
In this case, the energy density (which determines the number of
operations they can perform)
goes as the fourth
power of the temperature, and the entropy density (which determines the
number of bits they store)
goes as the
third power of the temperature.
It follows that
the total number of bits that can be registered using
conventional matter is just the number
of ops raised to the 3/4 power. Therefore, the number of bits
currently available is about
$(10^{123})^{3/4} \approx 10^{92}$.  And this yields an average separation
of about $10^{-3}$ cm between neighboring bits, which takes
light $10^{-14}$ sec to cross.

On the other hand, a computation rate of $10^{106}$ operations per sec
with $10^{92}$ bits means that each bit flips once every $10^{-14}$ sec.
So storing bits in ordinary matter
corresponds to a distribution of elementary events in which each bit flips
in the same amount of time it takes to communicate with a neighboring bit.
It follows that the accuracy to which ordinary
matter maps out spacetime corresponds exactly to the case of events spread out
uniformly in space and time discussed above, \cite{llo04, gio04}
(in the discussion above, replace clocks by bits and the events of the 
ticking of clocks by the flipping of bits),
or in other words, to the random-walk model of spacetime fluctuations.

To the extent that the random-walk model is observationally ruled
out, we infer that spacetime
can be mapped to a finer spatial resolution than
that given by $\delta l \sim l^{1/2} l_P^{1/2}$.  This in turn
suggests that there must be other kinds of matter or energy
with which
the universe can map out its spacetime geometry
(to a finer spatial accuracy than
that is possible with the use of conventional matter)
as it computes.
(Here we assume the validity of general relativity even at cosmological
scales.)
This line of reasoning\cite{critical}
strongly hints at, and thus provides an alternative argument for,
the existence of dark matter and
dark energy, independent of the evidence from recent cosmological
observations\cite{SNa,SNa1,WMAP,cluster}.

In conclusion, we
have suggested that we can test spacetime
foam models by looking for halo structures in the interferometric
fringes induced by fluctuations in the directions of the
wave vector of light from extragalactic sources.  This method is
considerably more powerful than tests based on spacetime foam-induced
phase incoherence alone.  (For the case of PKS1413+135 observed by HST, 
we obtain an improvement in sensitivity by 6 orders of magnitude.)
We have used it to rule out models of spacetime foam with
$\alpha \lesssim 0.6$, including the random-walk model.
On the other hand, the
holographic model predicts an angular spread in the spacetime foam
scattered halo which is still 3 orders of magnitude smaller than the
resolution of HST at $1.6 \mu$m.  But we note that the VLT and Keck
interferometers (plus future space interferometry missions such as SIM 
and TPF) may be on the verge of being able to test the
holographic model.  In particular, the full VLT
interferometer, when completed, will have four 8.2 m
telescopes with baselines up to $\sim 200$ m, plus four relocatable
$1.8$ m auxillary telescopes, which
may reach an angular resolution \cite{witt04} 
of 1-2 milliarcsec ($\sim 5$ to
$10 \times 10^{-9}$ radians) 
with a sensitivity of $\sim 50$ mJy
in the near infrared.  This is on the verge of testing
the holographic model, with a modest improvement in sensitivity and 
dynamic range. 

We also note that detection of
spacetime foam-induced phase fluctuations via interferometry requires
$\Delta \phi \sim l^{1-\alpha}/\lambda \sim
\lambda/D$.
Hence, for the purpose of testing
spacetime foam models, we can optimize the performance of the
interferometers by observing
light of shorter wavelengths from farther extragalactic point sources 
(subject to sensitivity limits) with
larger telescopes (like the recently proposed Extremely Large Telescope and
OverWhelmingly Large Telescope) of longer interferometer baseline.
It also may be noted that tests for spacetime foam effects can, in 
principle, be carried out without guaranteed time using archived high
resolution, deep imaging data on quasars and, possibly, supernovae
from existing and upcoming telescopes.

In this Letter, we have also argued that
the demise of the random-walk model (coupled with the fact that ordinary
matter maps out spactime only to the accuracy corresponding to the 
random-walk model)
can be used
to infer the existence of unconventional energy and/or matter.
This shows how spacetime foam physics
can even shed light on
cosmology\cite{likewise} --- a testimony to the unity of nature.

\bigskip

We thank L.L. Ng for his help in the preparation of the manuscript, 
and E.S. Perlman for his insightful comments and useful suggestions.
This work was supported in part by the Bahnson Fund at University
of North Carolina.


\begin{references}

\bibitem{Wheeler}
J.A. Wheeler, in
{\it Relativity, Groups and Topology}, edited by B.S. DeWitt and 
C.M. DeWitt
(Gordon \& Breach, New York, 1963), p. 315.

\bibitem{DeWitt}
B.S. DeWitt, Sci. Am. {\bf 249}, 112 (1983).

\bibitem{Hawking}
S.W. Hawking, D.N. Page and C.N. Pope, 
Nucl. Phys. {\bf 170}, 283 (1980).

\bibitem{Ashtekar}
A. Ashtekar, C. Rovelli and L. Smolin, 
Phys. Rev. Lett. {\bf 69}, 237 (1992).

\bibitem{Ellis}
J. Ellis, N. Mavromatos and D.V. Nanopoulos, 
Phys. Lett. B {\bf 293}, 37 (1992).

\bibitem{ng03b}
Y.J. Ng, 
Mod. Phys. Lett. A {\bf 18}, 1073 (2003).

\bibitem{AC}
G. Amelino-Camelia, 
Mod. Phys. Lett. A {\bf 17}, 899 (2002).

\bibitem{Turner}
L.M. Krauss and M.S. Turner, 
Gen. Rel. Grav. {\bf 31}, 1453 (1999).

\bibitem{SNa}
S. Perlmutter {\it et al.,}  
Astrophys. J. {\bf 517} 565 (1999).

\bibitem{SNa1}
A.G. Riess {\it et al.,}  
Astron. J. {\bf 116} 1009 (1998).

\bibitem{WMAP}
C.L. Bennett {\it et al.,}  
Astrophys. J. Suppl.
{\bf 148}, 1 (2003).

\bibitem{cluster}
S.W. Allen, R.W. Schmidt, A.C. Fabian and H. Ebeling, 
Mon. Not. Roy. Astron. Soc. {\bf 342},
287 (2003).

\bibitem{ng94}
Y.J. Ng and H. van Dam,
Mod. Phys. Lett.A {\bf 9}, 335 (1994).

\bibitem{ng95}
Y.J. Ng and H. van Dam,
Mod. Phys. Lett. A {\bf 10}, 2801 (1995).

\bibitem{kar66}
F. Karolyhazy, 
Il Nuovo Cimento A {\bf 42}, 390 (1966).

\bibitem{ng01}
Y.J. Ng,
Phys. Rev. Lett. {\bf 86}, 2946 (2001); (erratum) {\bf 88},
139902-1(E) (2002).

\bibitem{tho93}
G. 'tHooft, in
{\it
Salamfestschrift: a collection of talks}, edited by A. Ali,
J. Ellis, and S. Randjbar-Daemi
(World Scientific, Singapore, 1993) p. 284.

\bibitem{sus95}
L. Susskind,
J. Math. Phys. (N.Y.) {\bf 36}, 6377 (1995).

\bibitem{dio89}
L. Diosi and B. Lukacs,
Phys. Lett. A {\bf 142}, 331 (1989).

\bibitem{ame99}
G. Amelino-Camelia,
Nature (London) {\bf398}, 216 (1999).

\bibitem{giovanni}
In addition to Ref. \cite{ng94,ng95,kar66,ng01}, see also G. 
Amelino-Camelia, Mod. Phys. Lett. A{\bf 9}, 3415 (1994); Nature 
(London) {\bf 410}, 1065 (2001).

\bibitem{llo04}
S. Lloyd and Y.J. Ng, 
Sci. Am. {\bf 291}, 52 (2004).

\bibitem{gio04}
V. Giovannetti, S. Lloyd and L. Maccone, 
Science {\bf 306}, 1330 (2004).

\bibitem{mar98}
N. Margolus and L.B. Levitin, 
Physica D {\bf 120}, 188 (1998).

\bibitem{GPP}
It has been argued that the use of a relational notion of time and 
fundamental gravitational limits may slightly modify the Margolus-Levitin
theorem.  See R. Gambini, R.A. Porto and J. Pullin, 
arXiv:quant-ph/0507262.

\bibitem{average}
Note the qualification ``on the average.'' Shorter distances can be 
measured in one region but only at the expense of reduced precision
in some other regions.  This result is not inconsistent
with that found in X. Calmet, M. Graesser and S.D.H. Hsu, Phys.
Rev. Lett. {\bf 93}, 211101 (2004).

\bibitem{ng03a}
Y.J. Ng, W. Christiansen and H. van Dam, 
Astrophys. J. {\bf 591}, L87 (2003).

\bibitem{lie03}
R. Lieu and L.W. Hillman,
Astrophys. J. {\bf 585}, L77 (2003).

\bibitem{rag03}
R. Ragazzoni, M. Turatto and W. Gaessler,
Astrophys. J. {\bf 587}, L1 (2003).

\bibitem{per02}
E.S. Perlman {\it et al.,}  
Astron. J. {\bf 124}, 2401 (2002).

\bibitem{Handq}
Here we use $H_0 = 71$ km/s/Mpc and $q_0 = -0.5$.  Note that 
the luminosity distance $l$ is not appreciably changed even if 
we use, say, $q_0 = 0$, which we may incline to use if we have
not known of the accelerating cosmic expansion in the present 
and very recent cosmic eras.

\bibitem{cou03}
D.H. Coule,
Class. Quant. Grav. {\bf 20}, 3107 (2003).

\bibitem{explanation}
That the wave vector fluctuates with comparable magnitudes in
the transverse and longitudinal directions can be traced
intuitively to comparable
fluctuations in length in all directions. 

\bibitem{llo02}
S. Lloyd, 
Phys. Rev. Lett. {\bf 88}, 237901-1 (2002).

\bibitem{critical}
We note that a cosmic density an order of magnitude or so below
the critical density still would not qualitatively change the
conclusion.

\bibitem{witt04}
M. Wittkowski, 
Talk given in Workshop on ``The central parsec of galaxies''
at MPIA Heidelberg, 6-8 October 2004.

\bibitem{likewise}
Likewise, cosmology sheds light on spacetime foam physics.  That our
universe is at or very close to its critical density can be taken
as some sort of indirect evidence in favor of the holographic model 
($\alpha = {2 \over 3}$, corresponding to maximal spatial resolution).


\end{references}
\end{document}